# Observation of Collective Coulomb Blockade in a Gate-controlled Linear Quantum-dot Array

Wen-Yao Wei[1], Tung-Sheng Lo[2], Chiu-Chun Tang[1], Kuan-Ting Lin[3], Markus Brink[4], Dah-Chin Ling[5], Cheng-Chung Chi[1], Chung-Yu Mou[1], Jeng-Chung Chen[1, *], Dennis M. Newns[4], Chang C. Tsuei[4, *]

[1]Department of Physics, National Tsing Hua University, Hsinchu, 30013, Taiwan

[2] Institute of Physics, Academia Sinica, Taipei 115, Taiwan

[3]Institute of Industrial Science, The University of Tokyo, Komaba 4-6-1, Meguro-ku, Tokyo 153-8505, Japan

[4]IBM Thomas J. Watson Research Center, Yorktown Heights, NY 10598, U.S.A.

[5]Department of Physics, Tamkang University, Tamsui Dist., New Taipei City 25137, Taiwan

[*]Correspondence and requests for materials should be addressed to C. C. T. (tsuei@us.ibm.com) or J. C. C. (jcchen@phys.nthu.edu.tw).

The quantum transport of electrons in an artificial atom, such as a quantum dot (QD), is governed by the Coulomb blockade (CB) effects, revealing the ground-state charge configuration of the electronic system under interplays of the on-site strong Coulomb interactions[1]. In a coherently-coupled QD array, i.e. artificial molecules, the phenomenon of collective CB (CCB) was predicted by theoretical studies circa two decades ago but its evidence remains controversial[2,3]. Here, we present direct evidence for the observation of CCB in a six-quantum-dot array (QDA) under high magnetic fields at 20 mK. The coherent inter-dot coupling is enhanced and mediated via the Quantum Hall edge states of the GaAs sample substrate. Two continuously fine-tuned gate voltages enable the quantum dot conductance spectrum to undergo a localization to delocalization transition process which manifests as an emergence and a collapse of CCB. The transition between these two distinct quantum phases is analogous to the Mott-Hubbard metal-insulator transition[2,4,5]. The gate-voltage-dependent conductance map thus obtained makes it possible to utilize QDA as an on-chip laboratory for studying CCB and Mott physics[6-9]. Our QDA device provides a platform for developing engineered QD materials, qubit systems and artificial molecular devices in the future.

Semiconductor quantum dots (QDs) are man-made structures confining electrons within submicron islands[1]. Due to the nature of Coulomb interactions inherent, the energy spectrum of a QDs system can be theoretically casted into versatile quantum many-body systems, ranging from the Anderson Hamiltonian for a single QD[10,11] to the generalized Hubbard model for an array of QDs [2,6,8]. One of the outstanding examples to illustrate the resemblance of the many-body effects, predicted by the theory in 1994 but yet been experimentally confirmed[2,3], is that associated with increasing the inter-dot tunneling rate, the QDs system will undergo a correlated transport process from a state with collective Coulomb blockade (CCB) to a state with demolishing CCB. This transition can be analogous to the Mott-Hubbard metal-insulator transition, although it is not a true quantum phase transition for the finite size of the array[2,12,13]. The QDs system can mimic a Mott insulator in the presence of CCB, and a metal while CCB is collapsed. Experimentally, CCB could be featured as the emergence of grouped mini-CB conductance peaks period with a charging energy $U$ of a QD, resembled a band structure, while the breakdown of CCB manifests vanishing of $U$ or closing the band gap[2,8]. Tremendous experimental efforts have been made over years to disclose CCB, but a turnkey demonstration remains unaccomplished. There are two remaining major obstacles to significantly advance in simulating many-body physics with QDAs. First, despite using state-of-the-art nano-fabrication techniques, it is still difficult to implement a multi-dot system with a close quantum-dot spacing to achieve a substantial overlap of the outer shell wavefunctions. Second, the confining potential of the QDA formed by the gate voltages may vary from one dot to another, so the key issue is how to overcome this spatial fluctuation-induced disorder[8].

Here we adopt a novel approach to manipulate the inter-dot coupling via the assistance from the Quantum Hall edge states. In the presence of a perpendicular magnetic field, Integer

Quantum Hall effect (IQHE) takes place in the two dimensional electron gas (2DEG) and electrons propagate along conducting chiral edge channels[14]. The edge current possesses extremely long phase coherence due to the protection of the topological invariant[15]. The nature of the QH states in a single dot is well established[16, 17], but has not been applied to the study of a QDA. There are two primary advantages of using the IQH states in a multi-electron QDA. First, recent theories suggest that the screening effects in multi-electron qubits can effectively mitigate the potential fluctuations induced by the charged impurities[18, 19]. Second, in the presence of high magnetic fields, electrons in the dot can self-consistently adjust the confining potential by charge redistribution and the energy levels are aggregated[20, 21]. By utilizing these features, we have developed a means to equilibrate the variations of the dots in a QDA. We will present experimental results on a linear chain of 6-GaAs-based QDs to show that the edge-states-mediated interactions circumvent the effects of disorder, allowing us to provide the first evidence for collective Coulomb blockade (CCB) in conductance measurements.

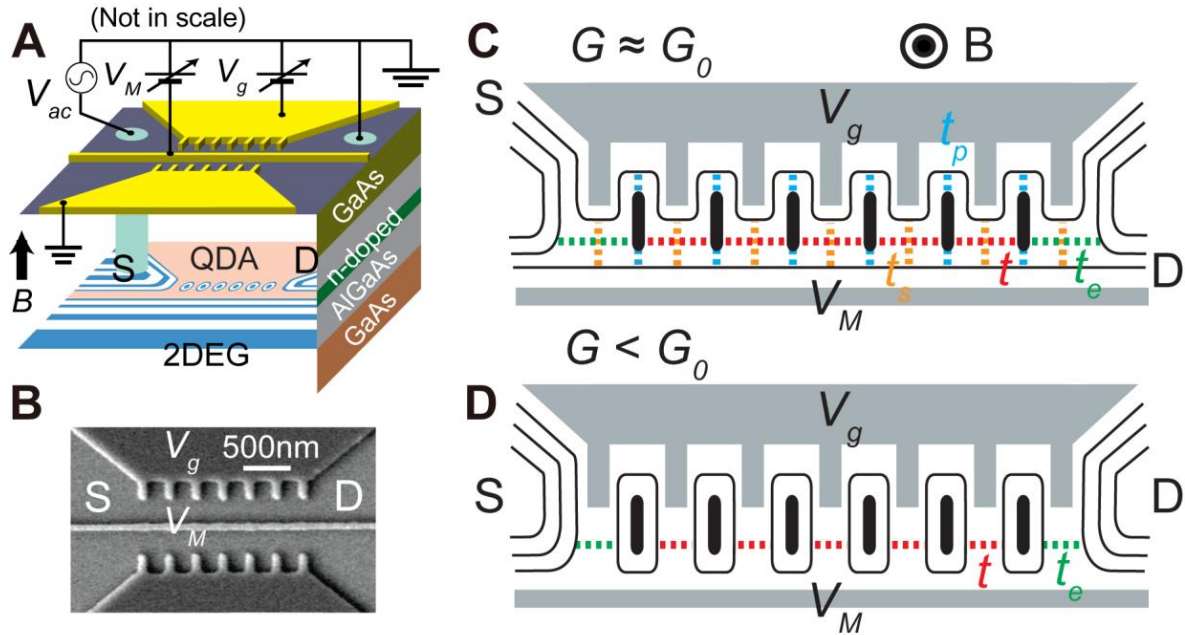

**Fig. 1.** Experimental setup and schematics to illustrate the electronic configurations of the QDA device in IQH regime. **(A)** The gated device structure, on top of a GaAs/AlGaAs heterostructure, consists of three gates: one upper finger gate with voltage $V_g$, one middle gate with voltage $V_M$, and one unused lower finger gate. The QDA is formed by confining the 2DEG to a chain of dots with the two end dots connected to source (S) and drain (D), respectively. **(B)** The SEM image of the device. **(C)** Schematic top view of the device to show the edge states in IQH regime. For $G \approx G_0$ ($= 2e^2/h$), one edge channel can flow along the boundary of the device and extend into the leads. **(D)** For $G < G_0$, all states are localized in the dot. The QDA consists of the metallic stadium-shaped core and outer ring[15].

The experimental arrangement is shown in Fig. 1A. The potential of the QDA can be adjusted by tuning $V_M$ or $V_g$, resulting in a change of electron number in each dot. Figure 1B shows scanning electron microscope (SEM) image of the device. In high magnetic field $B$, the electronic states of the QDA are driven into IQH states. Figures 1C and 1D illustrate the configurations of the edge channels for $G \approx G_0$ ($= 2e^2/h$) and $G < G_0$ regime, respectively. Four different tunneling paths are present[22]: (i) the inter-dot tunneling with amplitude $t$ (red horizontal lines), (ii) the tunneling between individual dot and the outer extended edge states with amplitude $t_p$ (blue vertical lines), (iii) the tunneling between the two end dots to the edge states in the source and drain leads with amplitude $t_e$ (green horizontal lines), and (iv) the tunneling between two opposite extended edge states with amplitude $t_s$ (orange vertical lines). In the case of $G \geq G_0$, the electronic states in the QDA consists of the localized and the extended edge states, marked by the black islands and lines in Fig. 1C, respectively. The extended states connect to the leads, whereas the localized states are formed into the QDA with metallic inner stadium-shaped cores and outer rings. The current-carrying extended edge states give rise to quantized conductance. Note that $G$ may deviate from the quantized value by tunneling electrons among the above-mentioned paths, which can be strongly modulated by Coulomb-blockade (CB) physics of the partially occupied Landau levels (LLs)[22]. For example, $t_p$ causes the backscattering between the extended edge channels and makes $G$ reduced. In contrast, $t_e$ connects two extra extended edge channels on the leads, yielding to an enhancement of $G$. For $G < G_0$, the channel width is narrowed down by applying less $V_g$ and $V_M$, and the extended edge states in the QDA disappear. The dots are serially coupled to each other, and charge transport is governed by $t$ and $t_e$. The QDA can therefore act as a quantum simulator for studying Mott physics. In this regime, if $V_M$ and $V_g$ are varied, we expect that $t$ and the number of electrons $N_e$ per dot change with modifying the confining potential of the QDA. These two control parameters – $N_e$ and $t$ – define the phase space in the studies of Mott transitions (see below).

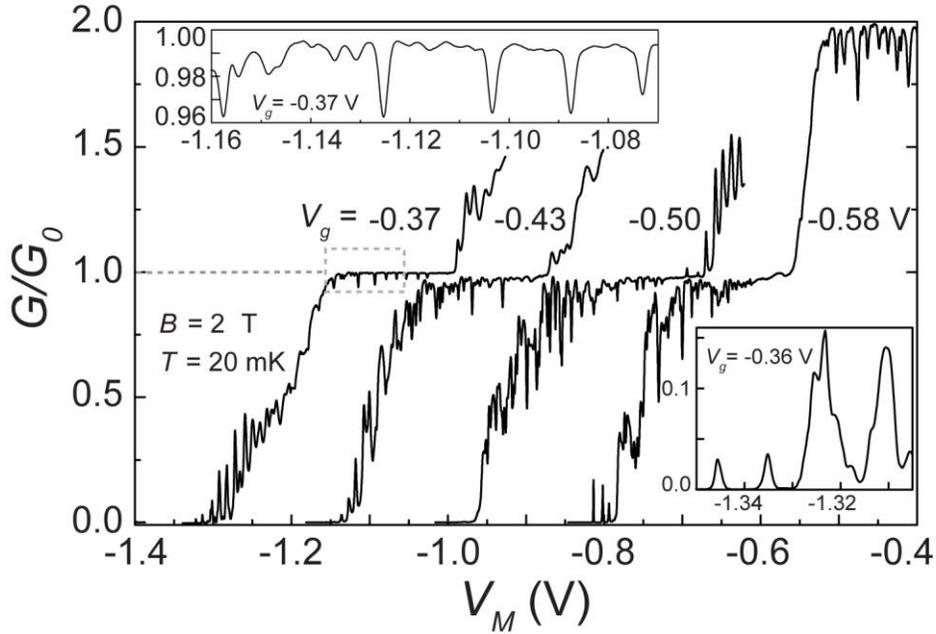

**Fig. 2.** Conductance characteristics of the QDA as a function of $V_M$ with various constant $V_g$ values. Conductance $G$ generically exhibits robust features, including quantized plateaus on $G_0$ and $2G_0$, the dip and peak structures superimposed on the plateaus, and the CB-type single and multiple peaks. The upper and lower insets show the representative traces to illustrate the detailed features on the $G_0$ plateau and in the CB regime, respectively.

To illustrate the entire conductance characteristics with voltage tuning, Fig. 2 shows the representative traces of $G$ versus $V_M$ with different $V_g$ at -0.37 V, -0.43 V, -0.50 V and -0.58 V. At $B = 2$ T, $G$ gradually increases to the quantized plateau at $G_0$ as $V_M$ is scanned. The $2G_0$ plateau is also visible at $V_g = $ -0.58 V. Note that these features and effects discussed below only observed at large magnetic fields, implying that Quantum Hall edge states are involved. Note that these features and effects discussed below only observed at large magnetic fields, implying that Quantum Hall edge states are involved. The observation of the quantized conductance strongly indicates: (i) the electron transport in the QDA device is essentially ballistic, thus the phase coherence of the carriers extends throughout the device[23,24], (ii) one or two fully transmitted edge channels flow through the array, as displayed in Fig. 1C, and (iii) the disorder effects do exist, but not dominate the transport properties. The hallmark of quantum ballistic transport in a quasi-one dimensional system is the appearance of quantized conductance plateaus in units of $G_0$. Each edge state in IQHE can be viewed as one transmitted channel and contributes a quantized plateau. One and two clear quantized plateaus are observed, as shown in Fig. 2, suggesting at least one or two transmitted ballistic channels are properly formed. Disorder can suppress the plateau features due to introducing extra backscattering processes. Nevertheless, it appears that the effects of disorder do not play a dominant role in our QDA device. Various dips and peaks superimposed on the $G_0$ plateau are observed under specific gating conditions and $B$. The upper inset of Fig. 2 shows an enlarged view of the dip structures, exhibiting mini-oscillations between two key deeper peaks. It has been suggested that the dip structures on the $e^2/h$ plateau in a QDA are related to the formation of energy bands in periodic potential barriers[3]. However, our data suggest that both the dip and peak structures arise from the dot-edge and the inter-edge tunnelings, i.e. the tunneling paths (ii) and (iii) as discussed earlier. In the regime of $G$

< $G_0$, the dots are tunneling-coupled in series and periodic CB-type $G$ peaks are observed. The lower inset of Fig. 2 reveals the emergence of multiple small peaks grouped into a broader peak, e.g. the peak at $V_M$ = -1.325 V, as a less negative $V_M$ is applied, namely the increase of $t$ (see below). The multiple peak features is the signal of the transport of electrons in a multi-QDs system[1].

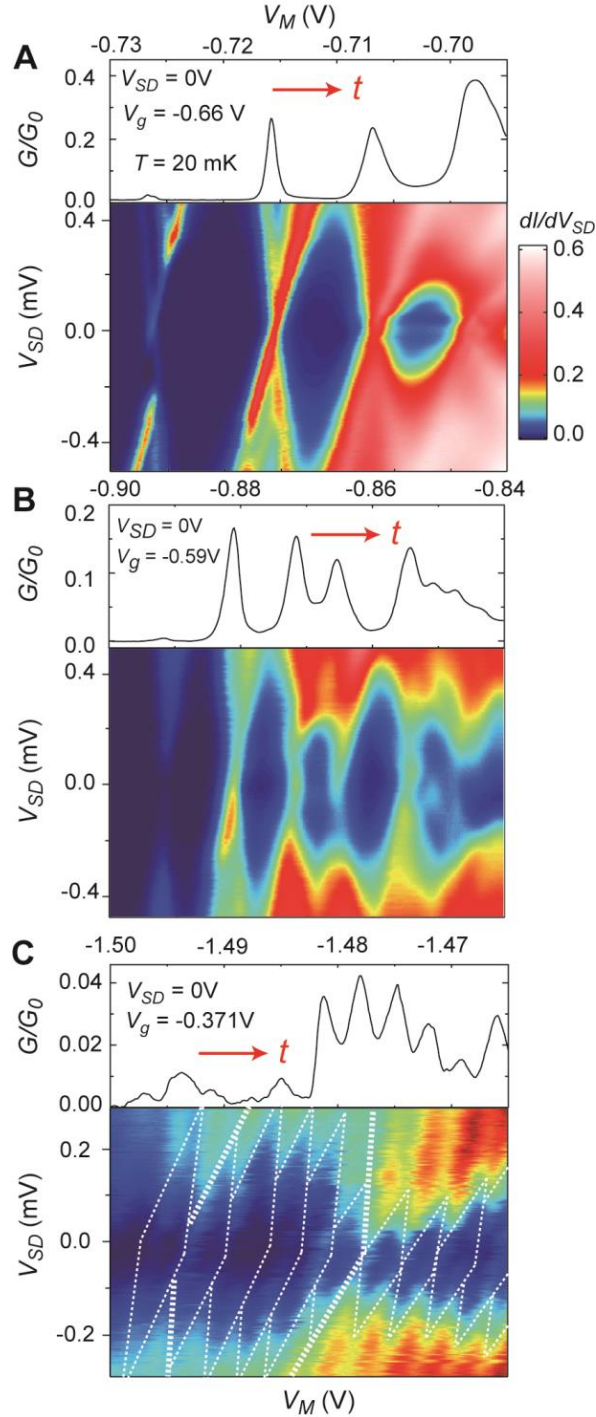

**Fig. 3.** The conductance peak profile below $G_0$ and the color-scale plot of differential conductance $dI/dV_{SD}$ as a function of $V_M$ and $V_{SD}$ in (A) weak $t$ regime, (B) intermediate $t$ regime, and (C) strong $t$ regime. The inter-dot coupling $t$ is tuned by adjusting $V_M$ and $V_g$. In general, $t$ is larger with more positive $V_g$ or $V_M$ (along the direction of the red arrow). The white dashed lines are marked to show the diamond structures for guides to the eye.

To further investigate the peak structure below $G_0$, we examine the evolution of the peak profile and the differential conductance $dI/dV_{SD}$ as a function of $V_M$ and $V_g$. Experimentally, $t$ increases with applying either less negative $V_g$ which makes the confining potential of QDA weaker or $V_M$ which pushes the dots towards the middle gate where the inter-dot barriers are smaller. Figure 3A to 3C show the $G$ traces (upper subfigure) and the associated $dI/dV_{SD}$ plot (lower subfigure) with increasing $t$ from weak to strong coupling regimes. Two distinct features disclose from the data set: (1) Two different kinds of peaks characterized by different $\Delta V_M$ period are observed: $\Delta V_M \sim 10$ mV (the single peak in Fig. 3A) in small $t$ and $\Delta V_M \sim 3$ mV (the mini- peaks in Fig. 3C) in large $t$. In the intermediate $t$ regime displayed in Fig. 3B, the mini-peaks are developed and then grouped into a broaden peak with the increase of $t$. (2) The $dI/dV_{SD}$ associated with the single peaks/ the mini-peaks reveals larger/smaller diamond structures in the $V_M$-$V_{SD}$ plot. The diamond structure unambiguously demonstrates that the observed conductance peaks originate from the CB effect. The current is blocked in the diamond-shaped areas shaded in dark blue where $N_e$ is fixed, whereas the current steps manifest as colored stripes located outside the diamond as $V_{SD}$ exceeds a threshold value. We can reasonably deduce that the larger $\Delta V_M$ (Fig. 3A) is dominated by the on-site charging energy $U$ (~1.5 meV) of the dot (Supplementary Material), and the mini-peak spacing (Fig. 3C) is in scale with the discrete level $\Delta$ (~0.3 meV) of a dot. We will show later that the mini-peaks originally are embedded in a single peak and can be gradually resolved on the broadening peak as $t$ increases. Features similar to ours have been observed in a multi-quantum-dot array[25] and in a single molecule transistor[26]. We hereby refer the single peak embedded with the mini-peak structure as collective Coulomb blockade[2,8] (CCB) peak to distinguish from the conventional CB peak in a single quantum dot. Each sharp and narrow peak corresponds to collectively adding multi-electrons to the array. The observation of CCB is possible only when the electrochemical potentials in the leads and the dots stay aligned, indicative of high uniformity of our QDA device. In terms of Mott physics, the energy gap leading to CCB is corresponding to the gap in a Mott insulator[2]. The observation of the pronounced CCB in the QDA implies that the potential variation possibly caused by charged impurities[9,18] is much less than the on-site Coulomb repulsion energy $U$. The collapse of CCB, accompanied by shrinking the width of the larger diamond and spreading the multiple peaks over the gap, with further increase of $t$ then can be viewed as a process analogous to Mott transition (MT) [2,8].

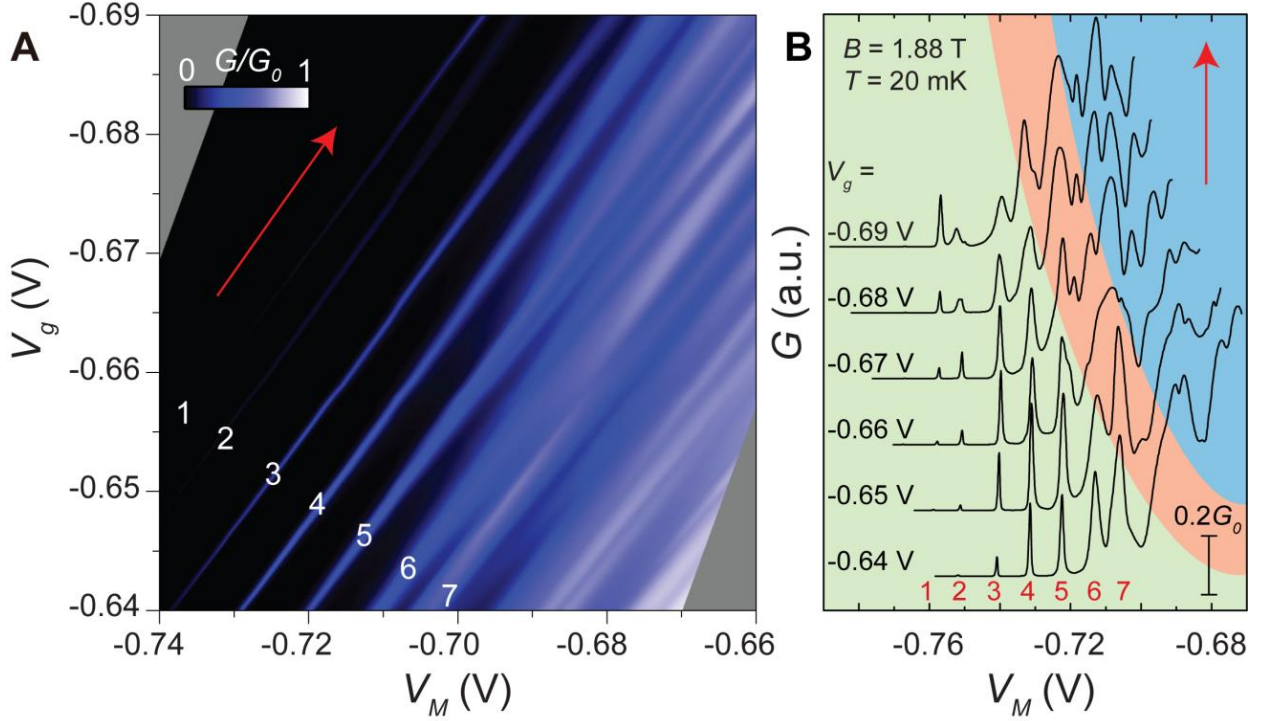

**Fig. 4.** **(A)** Conductance $G$ ($V_M$, $V_g$) demonstrates CCB (see text). Each peak in the weak $t$ regime is labeled by numbers for clarity. **(B)** To illustrate the evolution of the peak profile with $V_g$, selected traces from (A) with offsets in $V_M$ and $G$ are plotted. The CCB peaks, the collapsed CCB peaks, and the crossover regime are present in the green, blue, and red shaded areas, respectively.

We proceed by focusing on the continuous evolution of $G$ spectrum with different gating voltages to shed light on the underlying mechanism of the peak transitions. Figure 4A shows the $G$ ($V_g$, $V_M$) map of the QDA at $B = 1.88$ T. Narrow CCB peaks with $G \sim 0.2\ G_0$ are observed at the lower left corner of the map where $t$ is negligibly small. The QDA mimics as a Mott insulator in this area, denoted as the CCB regime, in which electrons are subjected to localization in each dot. As $t$ increases along the direction of the red arrow, while $N_e$ is fixed along the $G$ valley due to CCB, some of the CCB peaks such as peaks 6 and 7 exhibit a significant overlap. It turns out that new satellite peaks gradually develop during the merging of two peaks, which is associated with a delocalization of electrons from each individual dot, i.e. the breakdown of the CCB. The evolution of the peaks with increasing $t$, readily observed in Fig. 4B, imitates the Mott insulator to metal transition[2,8]. Multiple peaks in a strongly coupled multi-dot system correspond to the discrete levels inside the Hubbard bands (see the altering blue and white stripes in Fig. 4A and peaks in the blue shaded area in Fig. 4B). Adopting the approach in Ref. 2, we evaluate the crossover of two states occurs at $t/U \sim 0.1$ (Supplementary Material). Furthermore, it has been theoretically demonstrated that the detailed structure of the discrete levels within the Hubbard bands is sensitive to the presence of disorder[8]; consequently, the six mini-peaks associated with the six-QD chain are not expected to be clearly resolved due to an inevitable nonuniformity of the dots. However, the gross features of $G$ spectrum are robust against disorder[2].

In conclusion, we have revealed that the coherent inter-dot coupling is effectively mediated via the Quantum Hall edge states in our QDA device, which provides a Fermi level self-aligning mechanism to reduce the detrimental effect of disorder. It represents a ground-breaking advance for developing coherent 2D QDA devices and molecular artificial material despite the constraint of current micro-fabrication technology. We have clearly demonstrated that the emergence and breakdown of CCB, mimicking Mott-like transition, can be illustrated in a conductance map of a gate-controlled QDA. This work not only opens up an easy and new route to study many-body problems, but also provides a platform for engineering QD materials and artificial molecular devices.

**Acknowledgements** We acknowledge D. W. Wang, M. A. Cazalilla and L. M. K. Vandersypen for fruitful discussions and S. Komiyama for the experimental assistance. This project is supported by Ministry of Science and Technology Taiwan under grants NSC 101-2628-M-007-002-MY3, NSC 102-2112-M-032-005-MY3, and MOST 103-2112-M-007-016-MY3.